\documentclass[twocolumn]{jpsj3}

\usepackage{color}

\newcommand{\pdif}[2]{\frac{\partial #1}{\partial #2}}

\newcommand{\dakara}{.\raise1ex \hbox{.} . \, \, \,}
\newcommand{\nazenaraba}{\raise1ex \hbox{.} . \raise1ex \hbox{.} \, \, \,}

\newcommand{\expectation}[1]{\langle #1 \rangle}

\title{%
Mean-field Study of  
Charge, Spin, and Orbital 
Orderings \\
in Triangular-lattice Compounds $A$NiO$_2$ ($A$=Na, Li, Ag)
}

\author{%
Hiroshi Uchigaito\thanks{E-mail address: uchigaito@aion.t.u-tokyo.ac.jp}, Masafumi Udagawa, and Yukitoshi Motome
}

\inst{%
Department of Applied Physics, University of Tokyo, Hongo, Bunkyo-ku, Tokyo 113-8656, Japan
}

\abst{%
We present our theoretical results 
on the ground states 
in layered triangular-lattice compounds $A$NiO$_2$ ($A$=Na, Li, Ag).
To describe the interplay between charge, spin, orbital, and lattice degrees of freedom 
in these materials, 
we study a doubly-degenerate Hubbard model with electron-phonon couplings 
by the Hartree-Fock approximation combined with the adiabatic approximation.
In a weakly-correlated region, 
we find a metallic state  
accompanied by $\sqrt{3}\times\sqrt{3}$ charge ordering. 
On the other hand, 
we obtain an insulating phase with spin-ferro and orbital-ferro 
ordering 
in a wide range from intermediate to strong correlation.
These phases share many characteristics with the low-temperature states of AgNiO$_2$ and NaNiO$_2$, respectively. 
The charge-ordered metallic phase is stabilized 
by a compromise between 
Coulomb repulsions and effective attractive interactions originating from the breathing-type electron-phonon coupling as well as the Hund's-rule coupling. 
The spin-orbital-ordered insulating phase is stabilized by the cooperative effect of electron correlations and 
the Jahn-Teller coupling, 
while the Hund'-rule coupling also plays a role in the competition with other orbital-ordered phases. 
The results suggest a unified way of understanding a variety of low-temperature phases in $A$NiO$_2$. 
We also discuss 
a keen competition among different spin-orbital-ordered phases 
in relation to a puzzling behavior 
observed in LiNiO$_2$. }

\kword{%
multi-orbital Hubbard model, electron-phonon coupling, 
charge order, orbital order, metal-insulator transition, 
triangular lattice, 
NaNiO$_2$, LiNiO$_2$, AgNiO$_2$ 
}

\begin{document}
\maketitle

\section{Introduction}
\label{introduction}

One of the most distinctive features of strongly-correlated electron systems is diverse cooperative phenomena. 
~\cite{Imada,Tokura}
Correlated electron systems generally show intricate phase diagrams full of competing or coexisting states, and the phase competition often leads to  
exotic many-body phenomena.
Among many aspects, there are two factors which promote such complexity, i.e., 
multiple degrees of freedom
~\cite{Tokura,Kugel,Kugel2} 
and geometrical frustration~\cite{frustration}.
The multiple degrees of freedom are composed of 
charge, spin, and orbital of electrons.
Strong electron correlations induce interplay among them, resulting in a variety of phases. 
In addition, the interplay often causes  exotic response to external perturbations, such as the colossal magneto-resistance in perovskite manganites~\cite{Tokura2}.
On the other hand, the geometrical frustration promotes 
the phase competition.
In general, the geometrical frustration results in 
a huge number of low-energy degenerate states, 
by suppressing 
conventional long-range orders. 
The degeneracy yields nontrivial phenomena such as complicated ordering, glassy behavior, and spin-liquid states.
It is a long-standing problem in condensed-matter physics to understand a variety of phenomena 
which emerge from
synergetic effects between the multiple degrees of freedom and the geometrical frustration.

\begin{figure}[!t]
\begin{center}
\includegraphics[width=6cm]{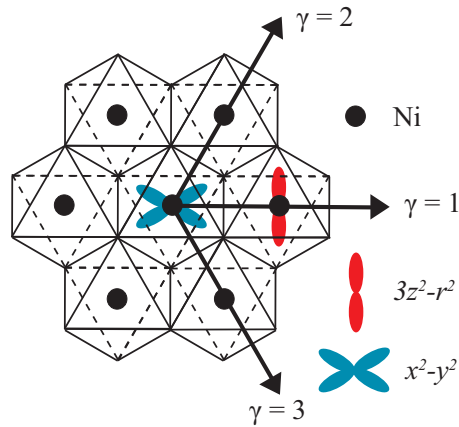}
\caption{(Color online). Schematic picture of NiO$_6$ layer in $A$NiO$_2$. NiO$_6$ octahedra share their edges with neighbors to form the triangular lattice of Ni cations. $\gamma$ denotes the Ni-Ni bond directions. 
Two $e_g$ orbitals, $3z^2 - r^2$ and $x^2 - y^2$, are shown.
}
\label{fig:triangular_layers}
\end{center}
\end{figure}

The family of compounds $A$NiO$_{2}$ ($A$=Li, Na, Ag) is a typical example of such geometrically-frustrated systems with multiple degrees of freedom.
$A$NiO$_2$ takes two different lattice structures depending on the cation $A$, that is, the delafossite structure~\cite{Sorgel2}
(AgNiO$_2$) and the ordered rock-salt structure~\cite{Chappel} (NaNiO$_2$ and LiNiO$_2$). 
Both structures are quasi-two-dimensional, composed of stacking of Ni, O, and $A$ layers.
The magnetic and transport properties are dominated by Ni cations, which are surrounded by the octahedron of oxygens.
The NiO$_6$ octahedra share their edges so that the Ni sites constitute the frustrated triangular layers as shown in Fig.~\ref{fig:triangular_layers}. 
Ni$^{3+}$ cation has seven $3d$ electrons in the low-spin configuration: Six out of seven 
fully occupy the lower 
$t_{2g}$ levels and the remaining one electron enters in the higher $e_g$ levels.
Hence the doubly-degenerate $e_g$ orbital degree of freedom is active in these systems.

$A$NiO$_2$ shows a variety of behaviors depending on the cation $A$ in spite of the similarity in lattice and electronic structure.
NaNiO$_2$ is a Mott insulator with a gap of 0.24eV~\cite{Molenda}. This compound shows a first-order structural transition accompanied by cooperative Jahn-Teller distortions at 480K~\cite{Chappel,Chappel2} and a second-order magnetic transition at 20K~\cite{Chappel2,Baker,Darie}. 
At the lowest temperature($T$), the system exhibits orbital-ferro and A-type 
antiferro-spin order (antiferromagnetic stacking of spin-ferro 
ordered layers)\cite{Lewis, Darie}. 
LiNiO$_2$ is also a Mott insulator with a gap of 0.2eV~\cite{Kanno}, however, it shows no clear phase transition down to 1.4K in contrast to NaNiO$_2$~\cite{LiNiO$_2$_noPT}. 
There is strong sample dependence in $T$ dependence of the
magnetic susceptibility; the field-cool and zero-field-cool  
bifurcation appears
in different ways depending on samples. This sample dependence strongly
suggests the relevance of extrinsic disorder~\cite{SG1}.
The ground state as well as the finite-$T$ 
properties remain controversial~\cite{SG1,SG2,SL1,SL2,Mostovoy,SR_ferromagneticcluster,muSR,Rougier,PDF}. 
In addition to these insulating materials, recently, a new metallic compound AgNiO$_2$ was synthesized~\cite{Sorgel2,Sorgel}. 
It shows a structural transition associated with a $\sqrt{3}\times\sqrt{3}$ charge ordering at 365K and antiferromagnetic transition at 20K~\cite{Wawrzynska,Wawrzynska2}. 
The system remains metallic down to the lowest $T$. 
It was claimed that in the low-$T$ phase the system separates into 
rather localized spins at 1/3 Ni sites and itinerant electrons at the remaining sites~\cite{Wawrzynska,Wawrzynska2}. 
Magnetic properties at low $T$ were analyzed by considering the competing 
nearest- and second-neighbor exchange couplings between localized spins~\cite{Wheeler,Coldea}. 

So far, $A$NiO$_2$ has been theoretically studied by the first-principle calculations and the strong-coupling analyses.
The orbital and spin ordering observed in NaNiO$_2$ was reproduced by the LSDA+$U$ first-principle calculations~\cite{Meskine,Meskine2}.
The metallicity as well as the charge ordering in AgNiO$_2$ was also reproduced by the first-principle calculations~\cite{Sorgel,Wawrzynska}.
On the other hand, effective models in the limit of strong electron correlation, the so-called Kugel-Khomskii models, were studied to understand the orbital and spin ordering in NaNiO$_2$ and the peculiar disordered state in LiNiO$_2$~\cite{Mostovoy,Dare,Vernay}. 
For AgNiO$_2$, recently, the magnetic phase diagram was investigated 
by the classical spin model with ignoring the itinerant electrons~\cite{Seabra}. 

Despite the extensive studies so far, comprehensive understanding of the ground states of $A$NiO$_2$ has not been reached yet.
Although the low-$T$ states of NaNiO$_2$ and AgNiO$_2$ are reproduced by the first-principle calculations, 
the mechanism of stabilizing these states is not fully clarified.
In addition, the effective-model approach does not fully succeed in reproducing the ground states of NaNiO$_2$ and LiNiO$_2$ in an unified way.
A possible way to explore the comprehensive understanding is 
to investigate a model  in a wide region of interaction parameters systematically 
beyond the strong-coupling approach. 
In fact, in both NaNiO$_2$ and LiNiO$_2$, the Mott gap is not large; 
the gap is comparable to the transfer integrals. 
Furthermore, the newly-synthesized AgNiO$_2$ shows metallic behavior. 
These facts suggest an importance of charge fluctuations 
in weakly- or intermediately-correlated regions. 
Electron-phonon couplings may be another essential factors, which 
have not been considered seriously in spite of the experimental facts that the structural transitions are observed in NaNiO$_2$ and AgNiO$_2$.

In this study, aiming at a unified picture of these compounds $A$NiO$_2$, we investigate 
a multi-orbital Hubbard model with electron-phonon couplings on a two-dimensional triangular lattice. 
Our purpose is to elucidate the microscopic mechanism for the variety of phases in these compounds $A$NiO$_2$.
In particular, we focus on 
electron-phonon couplings and charge degrees of freedom, both of which have not been 
carefully examined in the previous studies.
We clarify that these two elements play an important role in the phase competition in these complicated systems.
We obtain a $\sqrt{3}\times\sqrt{3}$ charge-ordered metallic (COM) phase in 
the weakly-correlated region 
and an insulating phase with ferro-type spin and orbital ordering in the intermediately- to strongly-correlated region.
These two phases reproduce many aspects of the low-$T$ states in AgNiO$_2$ and NaNiO$_2$, respectively.
We discuss the 
peculiar disordered state 
in LiNiO$_2$
in relation with a 
keen phase competition in the obtained 
phase diagram. 

The organization of this paper is as follows. In 
Sec.~\ref{model_and_method}, we 
describe our model and method. 
We introduce the Hamiltonian term by term, 
and present the approximations adopted in the calculations.
We 
present our results in 
Sec.~\ref{sec:results_and_discussion}. 
We discuss the parameter region and the mechanism to stabilize 
the $\sqrt{3}\times\sqrt{3}$ COM phase and 
the spin-orbital-ordered insulating phase.
Finally, 
Sec.~\ref{sec:conclusion} is devoted to summary. 

\section{Model and Method}
\label{model_and_method}

\subsection{Model}
In the present study, to 
elucidate the phase competition in $A$NiO$_2$, we investigate the ground state of the multi-orbital Hubbard model with electron-phonon couplings. Among the five $3d$ orbitals, we consider only the twofold degenerate $e_g$ orbitals, by 
taking account of the low-spin state of Ni$^{3+}$ cations. 
Our Hamiltonian is written as
\begin{equation}
\mathcal{H}= \mathcal{H}_{\rm{kin}}+\mathcal{H}_{\rm{int}}+\mathcal{H}_{\text{el-ph}} + \mathcal{H}_{\rm{ph}},
\label{Hamiltonian}
\end{equation}
where $\mathcal{H}_{\rm{kin}}$, $\mathcal{H}_{\rm{int}}$, $\mathcal{H}_{\text{el-ph}}$, and $\mathcal{H}_{\rm{ph}}$ represent the kinetic term of electrons, the electron-electron interactions, the electron-phonon couplings, and the elastic term of phonons, respectively. We describe the detailed forms of each term in the following.

\subsubsection{Kinetic term}
Due to the spatial anisotropy of the $e_g$-orbital wave functions, transfer integrals between Ni sites
depend on the bond direction as well as orbital types (see Fig.~\ref{fig:triangular_layers}). 
The kinetic term in eq.~(\ref{Hamiltonian}) is written as
\begin{equation}
\mathcal{H}_{\rm{kin}}  
= -\sum _{\expectation{ij}}\sum _{\alpha, \beta}\sum _{\sigma} t^{\gamma_{ij}}_{\alpha \beta} \bigl(c_{i\alpha \sigma}^{\dagger} c_{j\beta \sigma} +\rm{H.c}. \bigr).
\end{equation}
Here, $i$ and $j$ denote the site indices, $\alpha$ and $\beta$ represent the orbital indices with $\alpha=a(b)$ corresponding to the $3z^2-r^2$ ($x^2-y^2$) orbital, 
$\sigma$ is the spin, 
and $\gamma_{ij}$ denotes the direction of bond between the site $i$ and $j$, as
shown in Fig.~\ref{fig:triangular_layers}. 
The sum over $\expectation{ij}$ is taken for the nearest-neighbor sites on the triangular lattice.
The transfer integrals are given by the following matrices;
\begin{equation}
t^{\gamma=1} =
\begin{pmatrix}
t & \hspace{-4pt} 0 \\
0 & \hspace{-4pt} t^{\prime} 
\end{pmatrix}, 
\  
t^{\gamma=2} =
\begin{pmatrix}
\displaystyle
t_2 & \hspace{-4pt} t_3 \\
t_3 & \hspace{-4pt} t_4  
\end{pmatrix}, 
\ 
t^{\gamma=3} =
\begin{pmatrix}
t_2 & \hspace{-9pt} -t_3 \\
-t_3 & \hspace{-9pt} t_4 \label{eq:hopping_matrix_3}
\end{pmatrix},
\end{equation}
for the two bases of $3z^2-r^2$ and $x^2-y^2$ orbitals.
From the symmetry of orbitals, we obtain the following relations: $t_2=  t/4+3t'/4$, $t_3=\sqrt{3}(t-t^{\prime})/4$, and $t_4= 3t/4+t^{\prime}/4$, with
two independent parameters, $t$ and $t'$. 
We set $t=1$ as an energy scale. 
The value of $t'$ depends on both $d$-$d$ direct transfer integrals and 
$d$-$p$-$d$ indirect ones in a complicated manner~\cite{Dare,Vernay}. 
In the following, we show the results for $t'=-1$ 
by noting that the orbital overlaps between atomic orbitals at the neighboring sites 
lead to $t'/t \sim -1$ 
when one consider both contributions. 
An extended study in wider range of $t'/t$ for an effective model 
without phonon is found in Ref.~\citen{Vernay}. 
The choice of 
$t$ and $t'$ gives the non-interacting bandwidth $8t$.

\subsubsection{Electron-electron interactions} \label{sec:model_el-el}
Next we introduce the electron-electron interaction term $\mathcal{H}_{\rm{int}}$ in eq.~(\ref{Hamiltonian}). 
We consider only the on-site Coulomb interactions. 
For the doubly-degenerate $e_g$ orbital system, $\mathcal{H}_{\rm{int}}$ is written as
\begin{align}
\mathcal{H}_{\rm{int}} =   
\mathcal{H}_{U} + \mathcal{H}_{U'} + \mathcal{H}_{J_{\rm{H}}} + \mathcal{H}_{J_{\rm{H}}'}, 
\end{align}
where
\begin{align} 
\mathcal{H}_{U} = & \ 
U 
\sum_i \sum_\alpha n_{i\alpha \uparrow} n_{i\alpha \downarrow}, \\
\mathcal{H}_{U'} = & \ 
U^{\prime}  
\sum_i \sum_{\sigma\sigma'} n_{ia\sigma} n_{ib\sigma^{'}} , \\
\mathcal{H}_{J_{\rm{H}}} = & \ 
J_{\rm{H}} 
\sum_i \sum_{\sigma\sigma'}
c^{\dagger}_{i a \sigma} c^{\dagger}_{i b \sigma ^{\prime}}c_{i a \sigma^{\prime}} c_{i b \sigma}, \\
\mathcal{H}_{J_{\rm{H}}'} = & \ 
J_{\rm{H}}^{\prime}   
\sum_i \sum_{\alpha \neq \alpha'}
c^{\dagger}_{i \alpha \uparrow} c^{\dagger}_{i \alpha \downarrow} c_{i \alpha' \downarrow} c_{i \alpha' \uparrow} .
\end{align}
Here $n_{i\alpha \sigma}= c^{\dagger}_{i \alpha \sigma} c_{i \alpha \sigma}$, $U$ and $U^{\prime}$ denote the intra- and inter-orbital Coulomb repulsions, and 
$J_{\rm{H}}$ and $J_{\rm{H}}'$ denote the exchange interaction and the pair hopping, respectively. $\mathcal{H}_{J_{\rm{H}}}$ and $\mathcal{H}_{J_{\rm{H}}^\prime}$ are called 
the Hund's-rule couplings.
Hereafter we assume the relations 
$\ U'=U-2J_{\rm{H}}$ and $ J_{\rm{H}} = J_{\rm{H}} ^{\prime}$ 
to retain the rotational symmetry of the Coulomb interaction.

\subsubsection{Electron-phonon couplings} 

As to the electron-phonon couplings, we consider two relevant distortions of NiO$_6$ octahedra in $A$NiO$_2$, namely, the $A_{1g}$ breathing mode and the $E_g$ Jahn-Teller modes. $\mathcal{H}_{\text{el-ph}}$ in eq.~(\ref{Hamiltonian}) is given by the sum of these two contributions as 
\begin{equation}
\mathcal{H}_{\text{el-ph}}= \mathcal{H}_{\text{el-ph}}^{\rm{br}} + \mathcal{H}_{\text{el-ph}}^{\rm{JT}}.
\end{equation}

\begin{figure}[!t]
\begin{center}
\includegraphics[width=8cm]{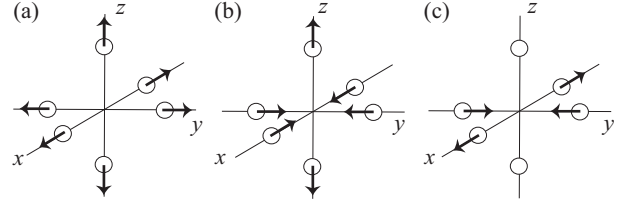}
\caption{Schematic pictures of the displacement of oxygens (open circles) around a Ni cation at the origin in (a) $A_{1g}$ mode, (b) $E_{g,u}$ mode, and (c) $E_{g,v}$ mode.}
\label{fig:A1g_Eg}
\end{center}
\end{figure}

The $A_{1g}$ mode corresponds to the isotropic expansion (contraction) of NiO$_6$ octahedron [Fig.~\ref{fig:A1g_Eg}(a)], which lowers (raises) two $e_g$ energy levels without lifting their degeneracy. Namely, the $A_{1g}$ mode couples to the 
local charge  
on each Ni site, written in the form
\begin{equation}
\mathcal{H}_{\text{el-ph}}^{\rm{br}} = -\gamma_{\rm{br}}  
\sum_i x_{{\rm br},i} \left(  n_{ia} + n_{ib} -1 \right), \label{eq:a1g}
\end{equation}
where $n_{i\alpha} = \sum_\sigma n_{i\alpha\sigma}$, 
$x_{{\rm br},i}$ is the amplitude of the $A_{1g}$ lattice distortion, and $\gamma_{\rm{br}} >0$ is the corresponding coupling constant. 
A positive (negative) $x_{{\rm br},i}$ corresponds to an expansion (contraction).

The $E_g$ mode has two components, 
$E_{g,u}$ and $E_{g,v}$, as shown in Figs.~\ref{fig:A1g_Eg}(b) and~\ref{fig:A1g_Eg}(c), respectively. The $E_{g,u}$ mode corresponds to the $z$-axis elongation of NiO$_6$ octahedron, which splits the energy levels of $x^2-y^2$ and $3z^2-r^2$ orbitals, while the $E_{g,v}$ mode causes a mixing of 
the two orbitals: 
The coupling to the $E_g$ modes is written as 
\begin{align}
\mathcal{H}_{\text{el-ph}}^{\rm{JT}} = 
-\gamma_{\rm{JT}} & \sum_{i,\sigma} \Bigl\{ x_{{\rm JT},i} \left( n_{ia\sigma}-n_{ib\sigma} \right) \notag \\
 & \quad + \bar{x}_{{\rm JT},i} ( c_{ia\sigma}^{\dagger} c_{ib\sigma}+c_{ib\sigma}^{\dagger} c_{ia\sigma} ) \Bigr\} , \label{eq:eg}
\end{align}
where $x_{{\rm JT},i}$ and $\bar{x}_{{\rm JT},i}$ are the amplitudes of $E_{g,u}$ and $E_{g,v}$ modes, respectively, and $\gamma_{\rm{JT}}$ represents the common coupling constant. 

\subsubsection{Phonon term}
The phonon term $\mathcal{H}_{\rm{ph}}$ in eq.~(\ref{Hamiltonian}) consists of the on-site term ($\mathcal{H}_{\rm{ph}}^{\rm{elastic}}$) and the inter-site term ($\mathcal{H}_{\rm{ph}}^{\rm{coop}}$) as,
\begin{align}
\mathcal{H}_{\rm{ph}} = & \mathcal{H}_{\rm{ph}}^{\rm{elastic}} + \mathcal{H}_{\rm{ph}}^{\rm{coop}}. \label{phonon}
\end{align}
Each term is given by the sum of contributions from the $A_{1g}$ and $E_g$ mode phonons.
The first term $\mathcal{H}_{\rm{ph}}^{\rm{elastic}}$ is the elastic energy of lattice distortions, which is given by the sum of the following two terms;
\begin{align}
\mathcal{H}_{\rm{ph}}^{\rm{elastic,br}} = & \frac{1}{2} \sum_i x_{{\rm br},i} ^2, \\
\mathcal{H}_{\rm{ph}}^{\rm{elastic,JT}} = & \frac{1}{2} \sum_i \left( x_{{\rm JT},i} ^2 + \bar{x}_{{\rm JT},i} ^2 \right).
\end{align} 
The elastic constants of $A_{1g}$ mode and $E_g$ modes  
are taken as unity without losing generality,
by normalizing the amplitudes of lattice distortions ($x_{{\rm br},i}$, $x_{{\rm JT},i}$, and $\bar{x}_{{\rm JT},i}$) 
and the coupling constants ($\gamma_{\rm{br}}$ and $\gamma_{\rm{JT}}$).

The second term $\mathcal{H}_{\rm{ph}}^{\rm{coop}}$ describes the cooperative couplings of lattice distortions.
The precise form of the couplings is, in principle, determined by the phonon dispersion in the materials, but here, for simplicity, 
we consider the nearest-neighbor couplings only.
Then $\mathcal{H}_{\rm{ph}}^{\rm{coop}}$ is defined by the sum of two terms;
\begin{align}
\mathcal{H}_{\rm{ph}} ^{\rm{coop,br}} = & \lambda_{\rm{br}} \sum _{\expectation{ij}}  x_{{\rm br},i} x_{{\rm br},j}, \\
\mathcal{H}_{\rm{ph}} ^{\rm{coop,JT}} = & - \lambda_{\rm{JT}} \sum _{\expectation{ij}} x_{{\rm JT},i} x_{{\rm JT},j} - \bar{\lambda}_{\rm{JT}} \sum _{\expectation{ij}}  \bar{x}_{{\rm JT},i} \bar{x}_{{\rm JT},j} , \label{eq:coop}
\end{align}
with the coupling constants $\lambda_{\rm{br}}$, $\lambda_{\rm{JT}}$, and $\bar{\lambda}_{\rm{JT}}$.
Although the values of $\lambda_{\rm{JT}}$ and $\bar{\lambda}_{\rm{JT}}$ are generally different, 
we take $\lambda_{\rm{JT}} = \bar{\lambda}_{\rm{JT}}$ for simplicity.
It is reasonable to assume $\lambda_{\rm{br}}$ to be positive, 
i.e., the ``antiferro"-type coupling, 
since the $A_{1g}$-type expansion (contraction) of NiO$_6$ octahedron tends to shrink (expand) the neighboring octahedra 
due to the edge-sharing network of octahedra. 
On the other hand, the tendency is opposite for the $E_g$ modes; 
the Jahn-Teller distortion of an octahedra induces the same distortion in the neighboring octahedra.
Hence we consider the ``ferro"-type coupling $\lambda_{\rm{JT}}>0$ in the following study.
In the following calculations, the ``antiferro"-type $A_{1g}$ coupling tends to 
stabilize a charge ordering by differentiating charge density between the neighboring sites, 
while the ``ferro"-type Jahn-Teller coupling favors ``ferro"-type orbital ordering.

\subsection{Method}

In order to study the ground state of the model (\ref{Hamiltonian}) 
in a wide range of parameters, we adopt the Hartree-Fock approximation to decouple the electron-electron interactions, and the adiabatic approximation to treat the electron-phonon 
couplings. 
Within the Hartree-Fock approximation, the two-body interaction terms 
in $\mathcal{H}_{\mathrm{int}}$ are decoupled by 
introducing mean fields, 
$\expectation{c_{i\alpha\sigma}^{\dagger} c_{i\alpha'\sigma'}}$. 
The amplitudes of lattice distortions are determined by the adiabatic approximation.
Within this approximation, the equilibrium values of $x_{{\rm br},i}$, $x_{{\rm JT},i}$, and $\bar{x}_{{\rm JT},i}$ are determined by using the Hellmann-Feynman theorem as
\begin{equation}
\Big\langle \pdif{\mathcal{H}}{x_{{\rm br},i}} \Big\rangle = 
\Big\langle \pdif{\mathcal{H}}{x_{{\rm JT},i}} \Big\rangle = 
\Big\langle \pdif{\mathcal{H}}{\bar{x}_{{\rm JT},i}} \Big\rangle = 0.
\end{equation}
These relations lead to the set of equations in the form
\begin{align}
x_{{\rm br},i}  + 
\lambda_{\rm{br}} {\sum_j }'  x_{{\rm br},j} = & \gamma_{\rm{br}} \left( \expectation{n_{ia} }+ \expectation{n_{ib} } - 1 \right), \label{eq:adi_br} \\
x_{{\rm JT},i} - 
\lambda_{\rm{JT}} {\sum _j }' x_{{\rm JT},j} = & \gamma_{\rm{JT}} \sum _{\sigma} \left( \expectation{n_{ia\sigma} }- \expectation{n_{ib\sigma} } \right), \label{eq:adi_JT} \\
\bar{x}_{{\rm JT},i}  - \lambda_{\rm{JT}} {\sum_j }' \bar{x}_{{\rm JT},j} 
= & \gamma_{\rm{JT}} \sum _{\sigma} 
( \expectation{ c_{ia\sigma}^{\dagger} c_{ib\sigma} } + \expectation{c_{ib\sigma}^{\dagger} c_{ia\sigma} } 
), \label{eq:adiabatic3}
\end{align}
where the sum $\sum '_j$ is taken over the nearest neighbors 
of the site $i$.

We determine the mean fields and the lattice distortions in a self-consistent way.
For a given set of 
$ 
\{
\expectation{c_{i\alpha\sigma}^{\dagger} c_{i\alpha'\sigma^{'}} }, x_{{\rm br},i}, x_{{\rm JT},i}, \bar{x}_{{\rm JT},i} \}$,  
we diagonalize the Hamiltonian under the Hartree-Fock approximation and obtain one-particle 
eigenenergies and eigenstates, 
which are used to calculate a new set of 
$ 
\{  
\expectation{c_{i\alpha\sigma}^{\dagger}  c_{i\alpha'\sigma^{'}} } \}$. 
These new mean fields are substituted in Eqs.~(\ref{eq:adi_br})-(\ref{eq:adiabatic3}) to determine the new set of 
$ 
\{ 
x_{{\rm br},i}$, $x_{{\rm JT},i}$, $\bar{x}_{{\rm JT},i} \}$. 
These procedures are repeated until the convergence is reached within the precision less than $%1.0 \times 
10^{-4}$ for all the variables.

\begin{figure}[!h]
\begin{center}
\includegraphics[width=8cm]{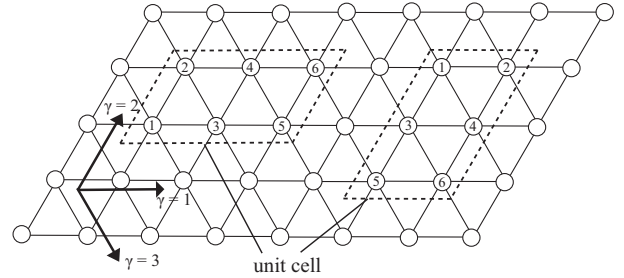}
\caption{
Two different ways of taking the 
unit cell with six Ni sites on the triangular lattice used in the calculations.
The axis $\gamma$ for the orbitals is shown (see also Fig.~\ref{fig:triangular_layers}).
}
\label{fig:unit_cell}
\end{center}
\end{figure}

In the following calculations, we take the unit cell which includes six Ni$^{3+}$ sites 
in the triangular lattice as shown in Fig.~\ref{fig:unit_cell}. 
To incorporate different orbital orderings depending on 
the way of taking the $\gamma$ direction in Fig.~\ref{fig:triangular_layers},
we consider two different ways of embedding the unit cell as shown in Fig.~\ref{fig:unit_cell}. 
These unit cells accommodate a $\sqrt{3}\times\sqrt{3}$ charge ordering and 
a 
two-sublattice ordering such as a stripe-type antiferromagnetic state.
Note that the ordering patterns observed in NaNiO$_2$ and
AgNiO$_2$ are both included by taking the unit cells. 
For the initial state in the iteration, we consider 
more than 30 different states with different symmetry in 
spin, orbital, and charge sectors, which are relevant in the parameter space we study.
The different initial configurations are adopted for 
each type of the unit cells in Fig.~\ref{fig:unit_cell}. 

The integration over the wave number in the calculations of the mean fields is replaced by the sum over $24 \times 24$ grids in the Brillouin zone for the supercell.
Hereafter, we focus on the quarter-filling case, i.e., 
one electron per site on average, 
corresponding to one $e_g$ electron in the low-spin state of Ni$^{3+}$.

\section{Results and Discussion}
\label{sec:results_and_discussion}

In this section, we 
show the results obtained for the ground state of the model~(\ref{Hamiltonian}).
In particular, we focus on the roles of $U$, $J_{\rm H}/U$, and $\gamma_{\rm{JT}}$, and discuss
how these parameters affect the ground state. 
We take $\gamma_{\rm{br}}=1.6$ 
and $\lambda_{\rm{br}}=\lambda_{\rm{JT}}=0.05$, 
which result in reasonable energy gain 
in forming CO, 
as we will see below. 

As a result, we find that the tendency to charge ordering 
becomes pronounced by a compromise among $U$, 
$U'$, $J_{\rm H}$, and $\gamma_{\rm br}$. 
On-site repulsions $U$ and $U'$ suppress charge 
disproportionation, while the Hund's-rule coupling $J_{\rm H}$ 
as well as the breathing-type coupling $\gamma_{\rm br}$ works as an inter-orbital 
effective attractive interaction 
and promotes charge disproportionation.
On the other hand, for larger $U$ and $U'$, 
the system becomes insulating, and   
the Jahn-Teller coupling 
$\gamma_{\rm{JT}}$ becomes important and 
enhances the tendency to orbital ordering, concomitant with magnetic ordering.
In order to characterize the orbital-ordered phases, we introduce the pseudospin operators in the orbital sector, defined as 
\begin{equation}
\boldsymbol{\tau}_{i} \equiv \sum _{\sigma} c^{\dag}_{i\alpha \sigma} \boldsymbol{\sigma}_{\alpha \beta} c_{i\beta \sigma}, 
\label{eq:pseudospin} 
\end{equation}
where $\boldsymbol{\sigma}_{\alpha \beta}$ denotes the pauli matrix.
For example, $\tau_z$-OF means a ferro-type order of $\tau_{iz}$, 
which is the $z$ component of $\boldsymbol{\tau}_i$.

\begin{figure}[!t]
\begin{center}
\includegraphics[width=6.5cm]{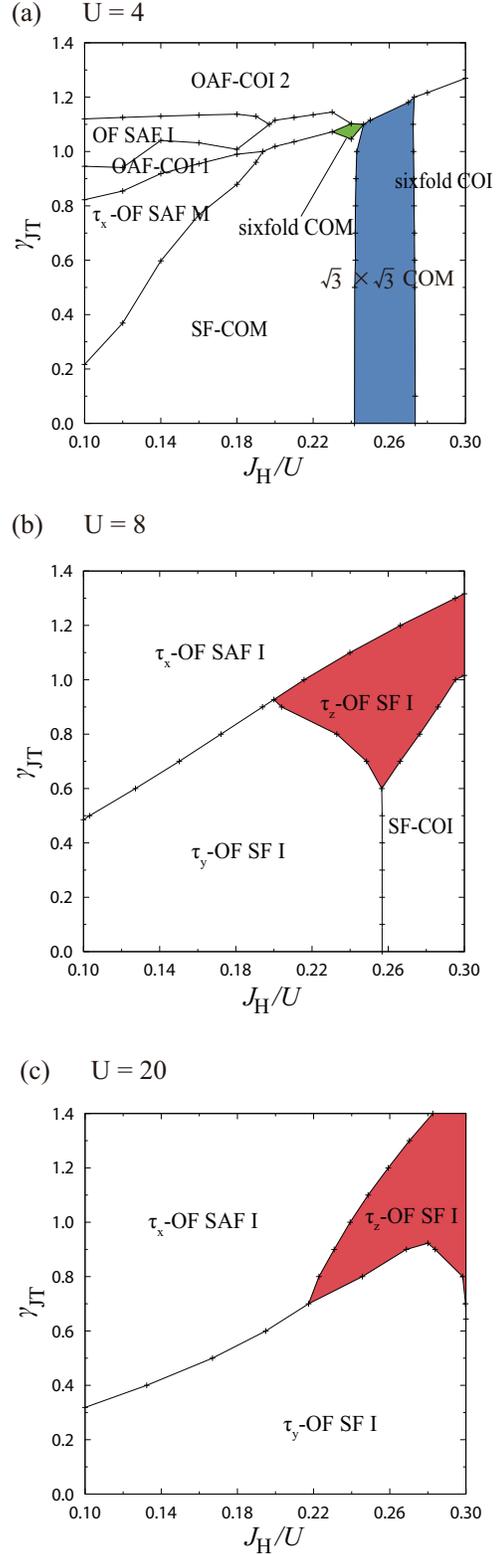}
\caption{(Color online).
Ground-state phase diagrams 
for the model (\ref{Hamiltonian}) 
in the plane of $J_{\rm H}/U$ and $\gamma_{\rm JT}$ 
at
(a) $U=4$, (b) $U=8$, and (c) $U=20$. 
We take $t'=-1$, $\gamma_{\rm br}=1.6$, and 
$\lambda_{\rm br} = \lambda_{\rm JT} = 0.05$. 
The ordering patterns of each phase are shown in Fig.~\ref{fig:sch}.
}
\label{fig:PD}
\end{center}
\end{figure}

The main result is summarized as the phase diagrams shown in Fig.~\ref{fig:PD}.
In the following, we will focus on the 
$\sqrt{3}\times\sqrt{3}$-type charger-ordered metallic (COM) phase found in the weak-coupling region [Fig.~\ref{fig:PD}(a)], and the
orbital-ferro 
spin-ferro 
insulating ($\tau_z$-OF SF I) phase obtained in the 
intermediately- to strongly-correlated region [Figs.~\ref{fig:PD}(b) and \ref{fig:PD}(c)].
These two phases are candidates for the low-$T$ states of AgNiO$_2$ and NaNiO$_2$, respectively. 
We will identify the parameter range for these phases, 
and discuss 
the origin and the nature of them in the following. 

\begin{figure}[!t]
\begin{center}
\includegraphics[width=7cm]{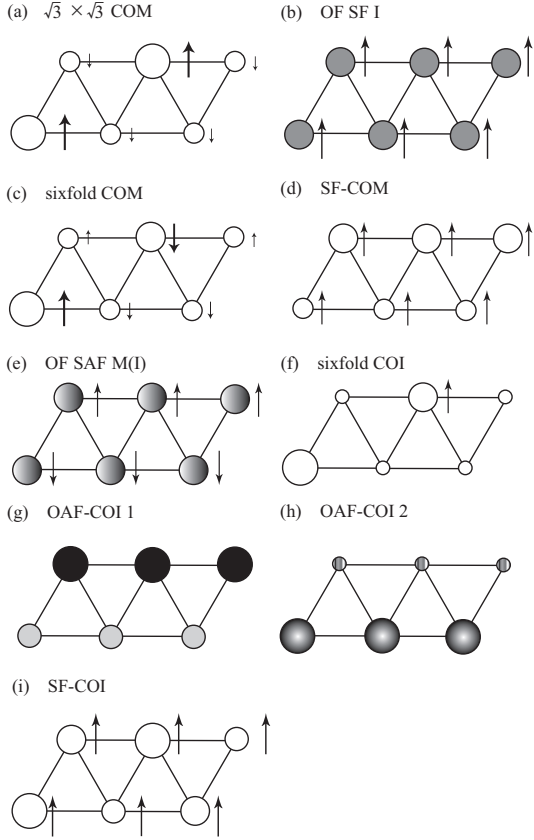}
\caption{
Schematic pictures of 
charge, spin, and orbital ordering patterns 
in the six-site unit cell
for the phases in Fig.~\ref{fig:PD}. 
The size of the circles schematically indicates the charge density at each Ni site.
The length of the arrows denotes the magnitude of the spin moment at each site.
The open circles denote the orbital para state, 
while the filled or shaded circles show orbitally polarized states.
The different patterns denote different orbitally polarized states.
See the text for details of (a)-(d). 
Both $\tau_z$-OF SF I and $\tau_y$-OF SF are represented by Fig.~\ref{fig:sch}(b) because these phases 
have the same symmetry (OF SF). 
Both OF SAF I and $\tau_x$-OF SAF M (I) are represented by Fig.~\ref{fig:sch}(e). 
Typical values of of the charge disproportionation and orbital polarization in (e)-(i) are as follows. 
(e) OF SAF I: $\expectation{\tau_x} \simeq \expectation{\tau_z} \simeq 0.6$ at all sites. 
$\tau_x$-OF SAF M: $\expectation{\tau_x} \simeq 0.2$ and $\expectation{\tau_z} \simeq -0.06$ at all sites.
(f): $\expectation{n} \simeq 3.0$ and $\expectation{n} \simeq 1.7$ at the two charge-rich sites and $\expectation{n} \simeq 0.3$ at the other charge-poor sites.
(g): $\expectation{\tau_x} \simeq 0.2$, $\expectation{\tau_z} \simeq 0.7$, and 
$\expectation{n} \simeq 1.2$ at the charge rich sites; 
$\expectation{\tau_x} \simeq 0.1 $, $\expectation{\tau_z} \simeq -0.2$, and $\expectation{n} \simeq 0.8$ at the charge poor sites.
(h): 
$\expectation{\tau_x} \simeq 1.4$, $\expectation{\tau_z} \simeq 0.6$, and $\expectation{n} \simeq 1.7$ at the charge rich sites and 
$\expectation{\tau_x} \simeq 0.2$, $\expectation{\tau_z} \simeq 0.0$, and $\expectation{n} \simeq 0.3$ at the charge poor sites.
(i): $\expectation{n} \simeq 1.02$ at the charge rich sites (1 and 4 in Fig.~\ref{fig:unit_cell}), 
$\expectation{n} \simeq 0.98$ at the charge poor sites (3 and 6), and 
$\expectation{n} \simeq 1.00$ at the other charge-neutral sites (2 and 5).
}
\label{fig:sch}
\end{center}
\end{figure}

\subsection{
Weakly-correlated region}
\subsubsection{Phase diagram}
A tendency toward charge ordering is widely observed 
in a weakly-correlated region. 
Figure~\ref{fig:PD}(a) shows the phase diagram at $U=4$. 
Among the competing phases, 
a COM phase 
with $\sqrt{3}\times\sqrt{3}$-type charge ordering
is stabilized 
in a wide range of $\gamma_{\rm JT}$ 
at $J_{\rm H}/U \sim 0.25$. 
The $\sqrt{3}\times\sqrt{3}$ charge ordering is
a three-sublattice order, in which one is charge rich (the density is  
almost two)
and the other two charge poor (the density is almost 0.5 per site).
This $\sqrt{3}\times\sqrt{3}$ COM state, shown in Fig.~\ref{fig:sch}(a), is remarkable, since 
it has the same 
charge ordering pattern 
as the low-$T$ state of AgNiO$_2$~\cite{Wawrzynska,Wawrzynska2}.
The spin state is also interesting; 
large moments appear at charge-rich sites ($S \simeq 0.6$), 
while the moments are suppressed at charge-poor sites ($S \simeq 0.05$). 
A similar differentiation was proposed for the low-$T$ state of AgNiO$_2$~\cite{Wawrzynska,Wawrzynska2}, 
although the calculated spin pattern does not fully agree with the experimental result.
We note that importance of interlayer coupling is experimentally suggested 
for the magnetic ordering~\cite{Wheeler,Wawrzynska,Wawrzynska2}, 
which is not taken into account in our model.
We will discuss the nature of this COM phase in Sec.~\ref{sec:COM_nature} in detail.

Around $J_{\rm{H}}/U=0.24$ and $\gamma_{\rm{JT}}=1.1$, we  
find another 
COM phase, 
i.e., the sixfold COM phase. Although this phase is stabilized in a narrow region in the phase diagram, it 
remains to be metastable in a wide parameter range, 
with a slightly 
higher energy 
than the ground state, as we will discuss 
in the next section \ref{sec:COM_stability}. 
This phase has the 
consistent ordering structure 
with the low-$T$ state of AgNiO$_2$ 
in terms of both 
charge and spin, as shown in Fig.~\ref{fig:sch}(c). 
Strictly speaking, the charge pattern of this phase has lower symmetry 
compared with AgNiO$_2$, due to the superposition 
of 
stripe-type charge 
modulation
onto the $\sqrt{3}\times\sqrt{3}$ charge ordering. However, 
the magnitude of 
this modulation
is very small: 
For charge rich sites (sites 1 and 4 in Fig.~\ref{fig:unit_cell}), we have $\expectation{n_1}=1.5040$ and $\expectation{n_4}=1.5027$, 
while for charge poor sites (sites 2, 3, 5, and 6 in Fig.~\ref{fig:unit_cell}), we have $\expectation{n_3}=\expectation{n_5}=0.7479$ and $\expectation{n_2}=\expectation{n_6}=0.7487$, 
at $J_{\rm H}/U=0.24$ and $\gamma_{\rm{JT}}=1.1$. 
The 
modulation gives 
very small charge disproportionation
of the order of $\sim 0.001$ within charge rich and poor sites. 
[The small disproportionations are exaggerated in the schematic picture in Fig.~\ref{fig:sch}(c).]

In addition to the two COM states, we obtain a variety of ordered phases in 
the weakly-correlated region.
Among them, we focus on 
two phases which compete with the COM states; 
the sixfold charge-ordered insulating (sixfold COI) phase in 
the large $J_{\rm H}/U$ region, 
and the 
spin-ferro metallic phase with a weak charge ordering 
(SF-COM) stabilized for smaller $J_{\rm H}/U$ [Fig.~\ref{fig:PD}(a)].
We argue the stability of the $\sqrt{3}\times\sqrt{3}$ COM  
as well as the sixfold COM in comparison with the two competing phases 
in the next section. 

\subsubsection{Stability of the 
$\sqrt{3}\times\sqrt{3}$ COM phase}
\label{sec:COM_stability}

\begin{figure}[!t]
\begin{center}
\includegraphics[width=8cm]{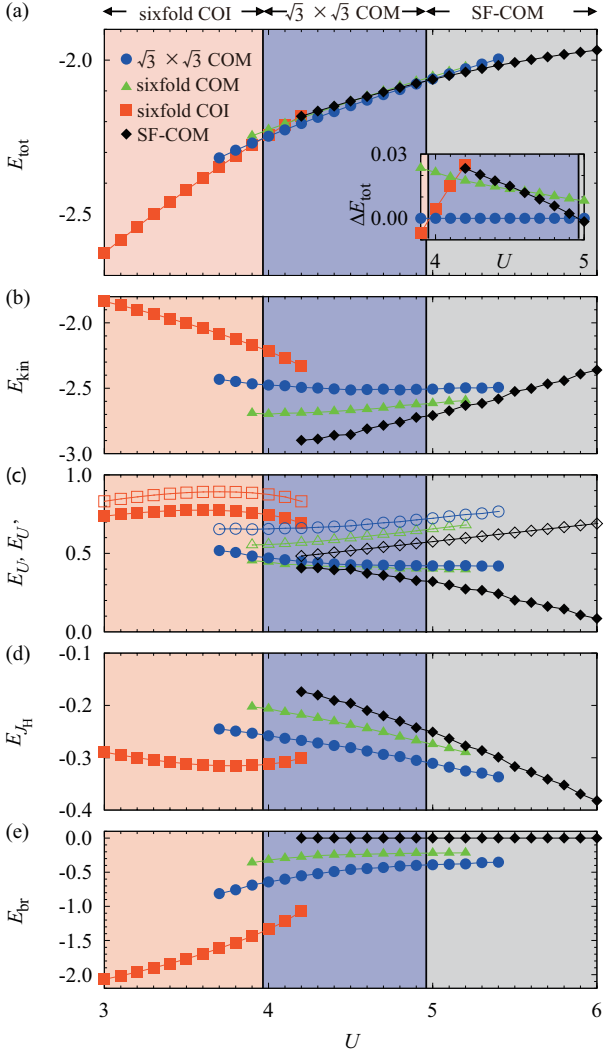}
\caption{
(Color online).
Energy comparisons among the $\sqrt{3}\times\sqrt{3}$ COM (circle), 
sixfold COM (triangle), sixfold COI (square), and SF-COM states (diamond): 
$U$ dependences of (a) the total energy, 
and the contribution from (b) kinetic term, 
(c) Coulomb repulsions [closed (open) symbols denote $E_U$ 
($E_{U'}$)],  
(d) Hund'-rule coupling, and (e) breathing-type electron-phonon coupling.
The inset of (a) shows the energy differences between the 
$\sqrt{3}\times\sqrt{3}$ COM phase and the other competing phases.
The parameters are $t'=-1$, $J_{\rm{H}}/U=0.27$, $\gamma_{\rm{br}}=1.6$, $\gamma_{\rm{JT}}=0.5$, and $\lambda_{\rm{br}}= \lambda_{\rm{JT}}=0.05$.
}
\label{fig:Uswp}
\end{center}
\end{figure}

In order to 
clarify the competition among the 
$\sqrt{3}\times\sqrt{3}$ COM, 
sixfold COM, sixfold COI, 
and 
SF-COM phases, we 
investigate the internal energy 
in detail by comparing the contributions from different terms in the Hamiltonian; 
$E_{\rm{kin}} \equiv \expectation{\mathcal{H}_{\rm{kin}}}$, $E_{U} \equiv \expectation{\mathcal{H}_{U}}$, 
$E_{U'} \equiv \expectation{\mathcal{H}_{U'}}$, 
$E_{J_{\rm{H}}} \equiv \expectation{ \mathcal{H}_{J_{\rm{H}}} + \mathcal{H}_{J_{\rm{H}}'} }$, 
$E_{\rm{br}} \equiv \expectation{ 
\mathcal{H}_{\text{el-ph}}^{\rm{br}} + \mathcal{H}_{\rm{ph}}^{\rm{elestic,br}} + \mathcal{H}_{\rm{ph}}^{\rm{coop,br}}}$, 
and the total energy
$E_{\rm{tot}} \equiv  \expectation{\mathcal{H}}$.
We show the comparison as a function of $U$ at $J_{\rm H}/U=0.27$ 
and $\gamma_{\rm JT} = 0.5$ in Fig.~\ref{fig:Uswp}. 

For $U \lesssim 3.9$, the sixfold COI 
state has the lowest energy. 
As shown in Fig.~\ref{fig:sch}(f), this phase has a polaronic nature, 
namely, one site is almost fully occupied (the local density is almost 4), 
and another one site accommodates almost two electrons. 
At the latter site, spins of two electrons are aligned parallel 
by the Hund's-rule coupling. 
This phase is stabilized in a region where the repulsive Coulomb interactions are 
compensated by effective attractive interactions originating in 
the breathing-type electron-phonon coupling as well as the inter-orbital Hund's-rule coupling. 
In fact, it is clearly observed in Figs.~\ref{fig:Uswp}(d) and \ref{fig:Uswp}(e)
that 
the energy gain in $E_{J_{\rm{H}}}$ and $E_{\rm{br}}$ 
contributes to the stabilization of the sixfold COI phase.

On the other hand, for $U \gtrsim 4.9$, the 
SF-COM state 
is most stabilized. 
As schematically shown in Fig.~\ref{fig:sch}(d), the charge ordering in this phase is 
a stripe type, but the charge disproportionation is very small 
($\expectation{n} \sim 1.03 - 1.09$ at charge rich sites, while $\expectation{n} \sim 0.91 - 0.97$ at charge poor sites): 
the main feature is the spin ferromagnetic ordering. 
The origin of this phase can be attributed to the Stoner mechanism~\cite{Stoner}.
As shown in the inset of Fig. \ref{fig:DOS}, the non-interacting Fermi level is located 
in the vicinity of the steep peak of the density of states (DOS).
Consequently, 
a ferromagnetic instability is caused at a relatively small $U \simeq 3.4$,
according to the Stoner's criterion. The characteristics of Stoner ferromagnet are 
observed in the
energy comparison in Fig.~\ref{fig:Uswp}(c); 
$E_U$ 
becomes smallest among 
the competing phases. 

The $\sqrt{3}\times\sqrt{3}$ COM 
state intervenes these two, 
and 
has the lowest energy for $4.0 \lesssim U \lesssim 4.8$. 
In the same parameter range, 
the sixfold COM 
state 
appears as a metastable state 
and 
stays very close to the ground state, as shown in the inset of Fig.~\ref{fig:Uswp}(a)  
(the energy difference is less than 0.02). 
These two COM states are stabilized by a compromise 
between the different stabilization mechanisms for the sixfold COI and the SF-COM phases. 
According to Fig.~\ref{fig:Uswp}, the COM phases have higher (lower) 
$E_{J_{\rm{H}}}$ and $E_{\rm{br}}$, while they have lower (higher) $E_U$ and $E_{U'}$, compared with
the sixfold COI phase (the
SF-COM phase). Namely, the two COM phases are stabilized in 
a subtle balance between 
repulsive Coulomb interactions and 
effective attractive interactions due to the Hund's-rule coupling and the breathing-type electron-phonon coupling.

Since 
the COM phases are stabilized in a delicate compromise,
it is important to consider 
their stability 
against the elements which are 
ignored in our current analysis, such as 
fluctuations beyond the mean-field level and the long-range part
of 
electron interactions. 
First, we 
note that the Stoner ferromagnetism is fragile 
when considering the electron correlation 
effect beyond the mean-field approximation\cite{Kanamori2}.
Therefore, the COM phases are expected to extend 
to larger $U$ or 
smaller $J_{\rm H}/U$. 
Second, the amplitudes of 
breathing-type distortions 
are fairly large in the sixfold COI phase compared to those in the other phases.
Hence this phase will be suppressed by considering 
more realistic contributions from phonons, e.g., 
anharmonic terms of phonons.
This may give a chance for the COM phases to become wider 
also in smaller $U$ or larger $J_{\rm H}/U$ regions. 
Finally, the long-range part of electron interactions generally works in favor of the charge ordering, in particular, the $\sqrt{3}\times\sqrt{3}$ type and the sixfold COM,
as is discussed in several transition metal compounds 
and organic materials~\cite{Seo}. 
Therefore, we expect that the COM phases relevant to AgNiO$_2$ 
become more stable in a wider parameter range 
when extending the analyses beyond the present model and method. 
Although the COM phases are robust in this parameter region, the energy difference between the sixfold COM phase and 
the $\sqrt{3} \times \sqrt{3}$ COM phase is 
very small, implying that the magnetic ordering pattern 
might be affected by 
small perturbations, such as inter-layer coupling.
More accurate studies 
beyond the mean-field approximation 
are necessary for 
fully determining the spin pattern of the ground state.

\subsubsection{Nature of the $\sqrt{3}\times\sqrt{3}$ COM phase}
\label{sec:COM_nature}

\begin{figure}[!t]
\begin{center}
\includegraphics[width=7cm]{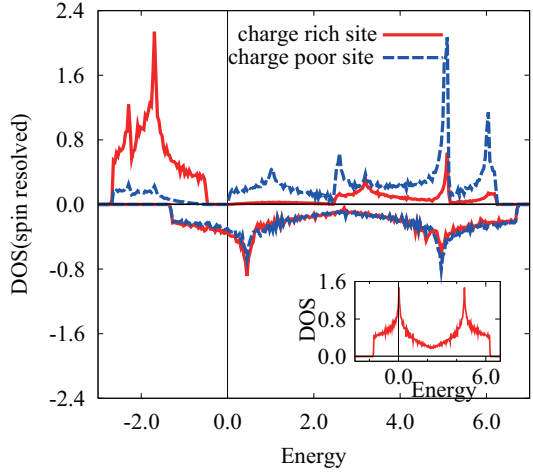}
\caption{
(Color online).
DOS per site
for the $\sqrt{3}\times\sqrt{3}$ COM phase. 
The Fermi level is set to be the origin of energy.
As to the $\sqrt{3}\times\sqrt{3}$ COM phase, the up-spin (down-spin) DOS is drawn on the upper (lower) side.
The parameters are chosen as 
$U=4$, $J_{\rm{H}}/U=0.27$, $\gamma_{\rm{br}}=1.6$, $\gamma_{\rm{JT}}=0.5$, and 
$\lambda_{\rm{br}}= \lambda_{\rm{JT}}=0.05$.
The inset shows DOS for the non-interacting case. 
}
\label{fig:DOS}
\end{center}
\end{figure}

Reflecting the subtle balance between the attractive and repulsive interactions, the $\sqrt{3}\times\sqrt{3}$ COM phase
shows 
peculiar electronic properties.
The density of states (DOS) in 
the $\sqrt{3}\times\sqrt{3}$ COM phase is shown in Fig.~\ref{fig:DOS}.
The site- and spin-resolved DOS 
indicates that the system exhibits a half-metallic 
nature: 
Up-spin electrons are localized at charge-rich sites, showing a gap at the Fermi level, 
on the other hand, 
down-spin electrons remain conductive, with a finite DOS at the Fermi level.
Electron correlations
dominantly affect 
up-spin electrons; 
down-spin conductive electrons preserve the non-interacting band structure.
For comparison, we show 
DOS for the non-interacting case in the inset of Fig.~\ref{fig:DOS}, which is
quite similar to DOS of down-spin conductive electrons.
DOS in the $\sqrt{3} \times \sqrt{3}$ COM phase reproduces several aspects of the result obtained by the first-principle band calculation~\cite{Wawrzynska}: The electrons at charge rich site tend to localize 
and the electronic structure at the other two charge poor sites resembles each other.

\begin{figure}[!t]
\begin{center}
\includegraphics[width=6cm]{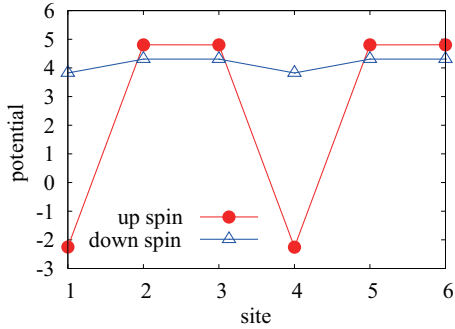}
\caption{
(Color online).
(a) 
Effective one-body potentials at each site for up-spin and down-spin electrons (see the text for details). 
The 
horizontal axis denotes the site indices in the unit cell, 
shown in Fig.~\ref{fig:unit_cell}.
The charge-rich (-poor) sites are the site 1 and 4 (2, 3, 5, and 6). 
The parameters are chosen as $U=4$, $J_{\rm{H}}/U=0.27$, $\gamma_{\rm{br}}=1.6$, $\gamma_{\rm{JT}}=0.5$, and 
$\lambda_{\rm{br}}= \lambda_{\rm{br}}=0.05$, consistent with Fig.~\ref{fig:DOS}.
}
\label{fig:potential}
\end{center}
\end{figure}

This peculiar electronic state can be attributed to the delicate balance between the
attractive and repulsive interactions. 
We plot the effective one-body potential in Fig.~\ref{fig:potential},
which is defined as 
the sum of 
the terms in
$\mathcal{H}_{\rm{int}}$ and $\mathcal{H}_{\text{el-ph}}$, which couple to the density operator $n_i$ at each site under the Hartree-Fock approximation.
Figure~\ref{fig:potential} shows that the charge-rich (-poor) sites 
bear attractive (repulsive) potentials 
for up-spin electrons. In contrast, the cancellation between the breathing-type electron-phonon coupling
and the repulsive Coulomb interactions leads to an almost flat potential for down-spin electrons.
Consequently, these interactions only work as a shift of chemical potential, and the down-spin
electrons retain the non-interacting band structure.

To conclude the discussions for the weakly-correlated region, 
the $\sqrt{3}\times\sqrt{3}$ COM phase is stabilized 
by a compromise between 
repulsive Coulomb interactions and 
attractive interactions 
originating from the breathing-type electron-phonon coupling 
as well as the Hund's-rule coupling.
This phase shows a half-metallic behavior with 
large magnetic moments almost localized at charge-rich sites 
and conduction electrons moving almost freely in the entire lattice.
We 
note that this phase is distinguished from the so-called pinball liquid state, in which the electrons at the charge-rich sites exclude the conduction electrons as hard core potentials and confine them to the honeycomb network of charge-poor sites, 
as discussed in a spinless tight-binding model with intersite Coulomb repulsion on the triangular lattice~\cite{pinball}.

\subsection{Intermediately- to strongly-correlated region}
Next, let us consider the intermediately- to strongly-correlated region. 
Representative phase diagrams are shown in 
Figs.~\ref{fig:PD}(b) and \ref{fig:PD}(c). 

\subsubsection{Phase diagram}
We show the ground-state phase diagram at $U=8$ in Fig.~\ref{fig:PD}(b).
Note that the value of $U$ is comparable with the non-interacting bandwidth., 
i.e., the system is in the intermediately-correlated region. 
In this region, we 
find three 
dominant orbital-ordered phases; 
$\tau_y$-ordered spin-ferro insulator ($\tau_y$-OF SF I), 
$\tau_x$-ordered spin-antiferro insulator ($\tau_x$-OF SAF I), and
$\tau_z$-ordered spin-ferro insulator ($\tau_z$-OF SF I).
The ordering pattern of $\tau_y$- or $\tau_z$-OF SF I ($\tau_x$-OF SAF I) is schematically shown in Fig.~\ref{fig:sch}(b) [Fig.~\ref{fig:sch}(e)].
For the small $\gamma_{\rm JT}$ and $J_{\rm H}/U$ region, 
$\tau_y$-OF SF I 
is stabilized, while 
it is replaced by 
$\tau_x$-OF SAF I
for larger $\gamma_{\rm JT}$
or by a charge-ordered state for larger $J_{\rm H}/U$. 
Meanwhile, 
when both $\gamma_{\rm JT}$ and $J_{\rm H}/U$ 
become large, 
$\tau_z$-OF SF I is stabilized. 
Among 
these phases, the 
$\tau_z$-OF SF I phase deserves attention, since 
the spin and orbital pattern of this phase is 
consistent with the low-$T$ 
phase of NaNiO$_2$.

Remarkably, 
the three OF phases remain stable in a wide range of $U$ toward the strongly-correlated regime. 
Figure~\ref{fig:PD}(c) shows an example of the phase diagram at large $U$. 
The result indicates that the three phases remain in similar parameter regions of 
$\gamma_{\rm JT}$ and $J_{\rm H}/U$ compared to the intermediate-$U$ case.
This
implies a possibility 
to understand the origin of these phases 
from the
strong-coupling analysis, i.e., by starting from the Mott insulating 
state at $U=\infty$. 

 In fact, 
 the $\tau_y$-OF SF I state was 
 obtained for an effective spin-orbital model in the strong-coupling limit 
(F3 phase of 
Fig. 8 in Ref.~\citen{Vernay}). 
Our result is consistent with the previous study. 
Meanwhile, the competition between the $\tau_x$-OF SAF I phase and the $\tau_z$-OF SF I phase is obtained for the first time
by explicitly taking account of 
electron-phonon couplings.
In the following, we will consider the mechanism of stabilization of these phases through the detailed study of energetics.

\subsubsection{Stability of the $\tau_z$-OF SF insulating phase}
\begin{figure}[!t]
\begin{center}
\includegraphics[width=8cm]{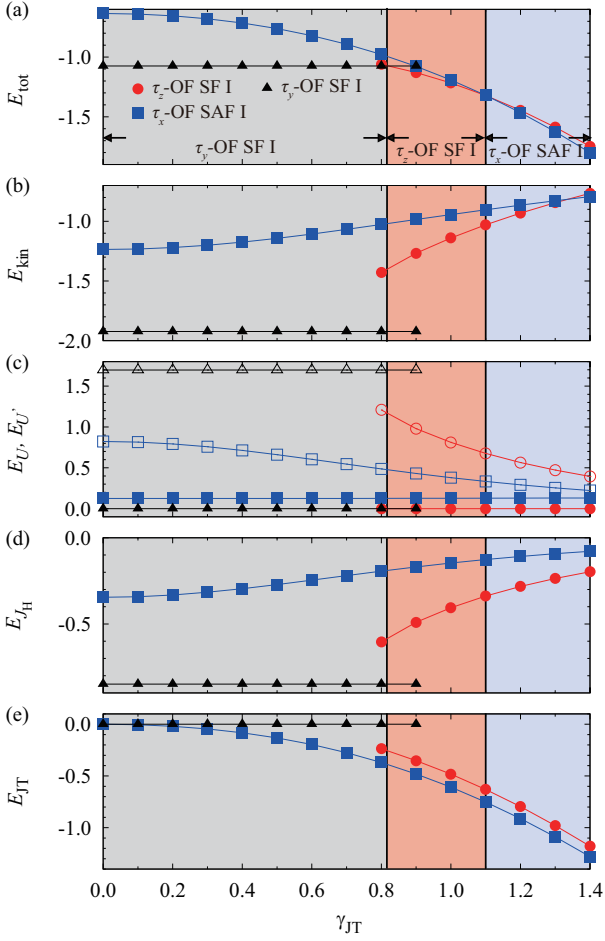}
\caption{
(Color online).
$\gamma_{\rm JT}$ dependences of 
(a) the total energy, and the contribution from 
(b) kinetic term, (c) Coulomb repulsions [closed (open) symbols denote $E_U$ ($E_{U'}$)], 
(d) Hund's-rule coupling, and (e) Jahn-Teller coupling. 
Comparison is made for the $\tau_z$-OF SF I (circle), 
$\tau_x$-OF SAF I (square), and $\tau_y$-OF SF I states (triangle).
The parameters are $t'=-1$, $U=20$, $J_{\rm{H}}/U=0.25$, $\gamma_{\rm{br}}=1.6$, and $\lambda_{\rm{br}} = \lambda_{\rm{JT}}=0.05$. }
\label{fig:JTswp}
\end{center}
\end{figure}

In order to understand the stability condition, it is instructive to rewrite $\mathcal{H}^{\rm{JT}}_{\text{el-ph}}$, $\mathcal{H}_{J_{\rm{H}}}$, and $\mathcal{H}_{J'_{\rm{H}}}$ 
by using the pseudospin operators in eq.~(\ref{eq:pseudospin}) as
\begin{equation}
\mathcal{H}_{\text{el-ph}}^{\rm{JT}} = -\gamma_{\rm{JT}} \sum_{i} 
( x_{{\rm JT},i}\tau_{iz} + \bar{x}_{{\rm JT},i}\tau_{ix} ), 
\label{tauexpression_JT}
\end{equation}
\begin{equation}
\mathcal{H}_{J_{\rm{H}}} + \mathcal{H}_{J_{\rm{H}}'} = \frac{J_{\rm H}}{2}\sum\limits_i 
\{  \tau_{ix}^2 - 
(n_{ia} 
 + n_{ib} 
) \}.
\label{tauexpression_JH}
\end{equation}
These equations clearly show that the Jahn-Teller coupling stabilizes the $\tau_x$ and $\tau_z$ orbital orderings, while the Hund's-rule coupling destabilizes the $\tau_x$ orbital ordering. 
In Fig.~\ref{fig:JTswp}, we 
compare the energy contributions including these terms, 
$E_{\rm{JT}} \equiv \expectation{ 
\mathcal{H}_{\text{el-ph}}^{\rm{JT}} + \mathcal{H}_{\rm{ph}}^{\rm{elastic,JT}} + \mathcal{H}_{\rm{ph}}^{\rm{coop,JT}}}$ and $E_{J_{\rm H}}$, 
together with other relevant energy contributions, 
for the three 
orbital-ordered phases. 

Figure~\ref{fig:JTswp}(b) shows that the stability of 
the $\tau_y$-OF SF I phase 
in the 
small $\gamma_{\rm{JT}}$ region 
is attributed to 
the energy gain in the kinetic energy $E_{\rm{kin}}$.
This is consistent with the result of strong-coupling analysis, where the kinetic energy gain through the spin-orbital superexchange interactions is claimed to
be the origin of this phase.~\cite{Vernay} 
It is also observed that all energy contributions in this phase 
are fairly insensitive to $\gamma_{\rm JT}$, as shown in Fig.~\ref{fig:JTswp}. 
This is expected from the absence of coupling between 
$\tau_y$ 
and 
Jahn-Teller distortions, 
as is clear from eq. (\ref{tauexpression_JT}).

On the other hand, the $\tau_x$-OF SAF I and 
$\tau_z$-OF SF I states lower their energy through the coupling to 
the Jahn-Teller distortions [Fig.~\ref{fig:JTswp}(e)], 
as expected from eq. (\ref{tauexpression_JT}); thus
they replace the $\tau_y$-OF SF I phase 
and become 
the ground state for larger $\gamma_{\rm JT}$, as shown in Fig.~\ref{fig:JTswp}(a).
The Hund's-rule coupling plays an important role in the relative stability between the $\tau_x$-OF SAF I and the $\tau_z$-OF SF I phases. 
As is evident from eq. (\ref{tauexpression_JH}), the Hund's-rule coupling affects only the $\tau_x$ ordering, and destabilizes it [Fig.~\ref{fig:JTswp}(d)]. 
Furthermore, the $\tau_z$-OF SF I phase is stabilized by the kinetic energy gain 
from the interorbital hoppings, compared with the $\tau_x$-OF SAF I phase as shown in Fig.~\ref{fig:JTswp}(b).
In fact, 
according to the second-order perturbation from the strong coupling limit $U \to \infty$, the $\tau_z$-OF SF I phase has lower energy than $\tau_x$-OF SAF I phase 
for $J_{\rm H}/U > \frac{8-\sqrt{10}}{18} \simeq 0.27$. 
The phase boundary between the two phases in Fig.~\ref{fig:PD}(c) is roughly located 
around this critical value, which indicates that 
the phase competition is essentially understood from the strong coupling picture.

\subsubsection{Nature of the $\tau_z$-OF SF insulating phase}

\begin{figure}[!t]
\begin{center}
\includegraphics[width=8cm]{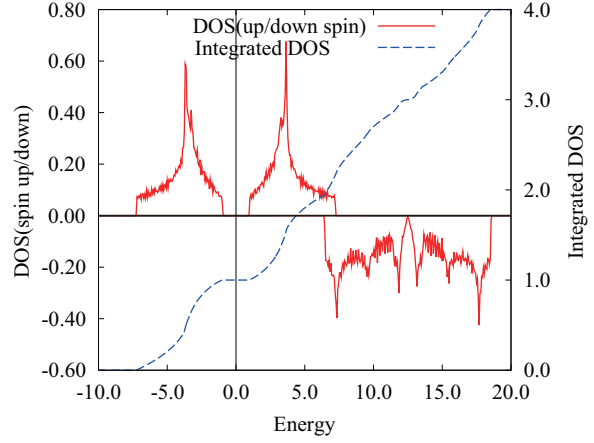}
\caption{
(Color online). 
DOS 
for the $\tau_z$-OF SF I 
state. 
The parameters are $t'=-1.0$, $U=20$, $J_{\rm{H}}/U=0.25$, $\gamma_{\rm{JT}}=1.0$, $\gamma_{\rm{br}}=1.6$,  and $\lambda_{\rm{br}}=\lambda_{\rm{JT}}=0.05$. The vertical line denotes the Fermi level. The up-spin (down-spin) DOS is represented on the upper (lower) side.
The integrated DOS is also shown.
}
\label{fig:dos_Na}
\end{center}
\end{figure}

Due to the strong electron repulsion, an excitation gap opens at the Fermi level in the $\tau_z$-OF SF I phase.
Hence, this state is insulating, consistent with the low-$T$ 
insulating phase of NaNiO$_2$.
Figure~\ref{fig:dos_Na} shows DOS in the $\tau_z$-OF SF I phase.
DOS is composed of four sectors  (two in each spin component),  
and the total weight of each sector is equal to one electron per site, 
as indicated in the integrated DOS in the figure.
This structure can be qualitatively understood by considering the atomic limit with ignoring the hoppings, $t$ and $t'$. 
Let us assume the perfect $\tau_z$-OF SF order, i.e., $\langle n_{i,3z^2-r^2,\uparrow}\rangle=1$ and let other mean fields to be zero. 
Then the excitation energies in the atomic limit are estimated as
\begin{align}
E_{x^2-y^2,\uparrow} &= U' -J_{\rm{H}} -2E^{*}_{\rm{JT}},\label{eq:gap2}\\
E_{x^2-y^2,\downarrow} &= U' -2E^{*}_{\rm{JT}},\label{eq:gap3}\\
E_{3z^2-r^2,\downarrow} &= U +2E^{*}_{\rm{JT}}, \label{eq:gap4}
\end{align}
where $E_{\alpha\sigma}$ signifies the energy necessary to add one electron with 
orbital $\alpha$ and spin $\sigma$
to the $(3z^2-r^2,\uparrow)$ ground state.
$E^{*}_{\rm{JT}}$ is the 
energy gain from the 
Jahn-Teller distortion per site, estimated as 
\begin{equation}
E^{*}_{\rm{JT}}= -\frac{ \gamma _{\rm{JT}} ^2}{2(1-6 \lambda _{\rm{JT}})}. %$.
\end{equation}
Substituting the parameters 
used in Fig.~\ref{fig:dos_Na} ($U=20$, $U'=10$, $J_{{\rm H}}=5$, $\gamma_{\rm JT}=1.0$, and $\lambda_{\rm JT}=0.05$) into 
these equations, we 
obtain $E_{x^2-y^2,\uparrow}\simeq6.4$, $E_{x^2-y^2,\downarrow}\simeq11$, and $E_{3z^2-r^2,\downarrow}\simeq19$. %, 
These values well correspond to the mean energy of each sector of DOS 
measured from that for the lowest one in Fig.~\ref{fig:dos_Na}. 

\begin{figure}[!t]
\begin{center}
\includegraphics[width=7cm]{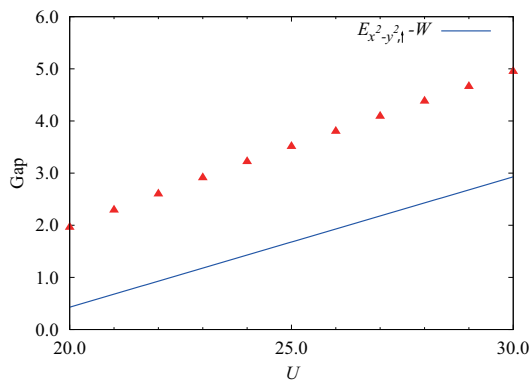}
\caption{
(Color online).
Energy gap in the $\tau_z$-OF SF 
I phase as a function of $U$ at $t'=-1.0$, $J_{\rm{H}}/U=0.25$, $\gamma_{\rm{JT}}=1.0$, $\gamma_{\rm{br}}=1.6$, and $\lambda_{\rm{br}}=\lambda_{\rm{br}}=0.05$. 
The 
line is the simple estimate from 
the strong-coupling analysis. 
See the text for details.}
\label{fig:gap}
\end{center}
\end{figure}

Eq.~(\ref{eq:gap2}) gives an estimate of the energy gap $\Delta$ in the atomic limit.
This atomic value is reduced for finite $t$ and $t'$, 
since 
each atomic level is broadened by
a renormalized bandwidth $\tilde{W}$: 
The estimate of energy gap is modified as $\Delta\sim E_{x^2-y^2, \uparrow} -\tilde{W}$.
We 
use this simple estimate 
with replacing $\tilde{W}$ by the bare bandwidth $W=8t$
for a comparison 
with 
the Hartree-Fock solutions. 
As shown in Fig.~\ref{fig:gap}, our simple estimate
from the atomic limit is qualitatively consistent with 
Hartree-Fock results. 
This fact 
also supports that the electronic spectrum of the $\tau_z$-OF SF I phase is adiabatically continued from 
the strong-coupling limit. 


To conclude this part, three orbital-ordered insulating phases appear dominantly 
in the intermediately- to strongly-correlated region. 
Among them, the $\tau_z$-OF SF I phase, which is relevant to NaNiO$_2$, 
becomes stable in the region where 
both the Jahn-Teller coupling and the Hund'-rule coupling are substantial. 
This phase is understood by the strong-coupling picture under the 
Jahn-Teller type electron-phonon couplings.

\subsection{Comparison to experiments} \label{sec:comparison}

In our calculations, we successfully reproduce the $\sqrt{3}\times\sqrt{3}$ COM phase and 
the $\tau_z$-OF SF 
I phase, 
whose ordering patterns are consistent with 
the low-$T$ phases in AgNiO$_2$ and NaNiO$_2$, respectively.
These two phases appear in close parameter regions of 
$\gamma_{\rm JT}$ and $J_{\rm H}/U$, 
but for different range of 
the on-site repulsion $U$. 
The $\sqrt{3}\times\sqrt{3}$ COM phase is stabilized 
in the weakly-correlated region, 
where $U$ is smaller than the bare bandwidth, 
while the $\tau_z$-OF SF 
I phase 
is stabilized in 
the intermediately- to strongly-correlated region. 

The difference in $U$ 
may be attributed to the structure of cation bands in these compounds.
The magnitude of the effective on-site repulsion 
for Ni $3d$ electrons 
is not solely determined by its atomic value, but it is considerably affected by the
screening effect brought about by 
electrons in the $A$ cation and oxygen $p$ bands.
According to the first-principle calculations, 
Ag bands in AgNiO$_2$ reside in the vicinity of the Fermi level~\cite{Sorgel,Wawrzynska}, 
while the Na bands in NaNiO$_2$ are located about 4eV above the Fermi level~\cite{Meskine,Meskine2}.
Consequently, a larger screening effect is expected for AgNiO$_2$, which reduces the magnitude of $U$ considerably, 
compared to that for NaNiO$_2$.
Our results are consistent with this trend. 
In the first-principle calculations~\cite{Meskine,Meskine2,Sorgel,Wawrzynska}, the bandwidth of 
Ni 3$d$ bands was roughly estimated to be $2\sim3$eV, 
leading to a rough estimate of $t=0.25\sim0.4$eV.
In the LSDA+$U$ calculation for NaNiO$_2$~\cite{Meskine,Meskine2}, the value of $U$ was taken to be 5eV to reproduce the correct size of excitation gap.
On the other hand, a cluster-model analysis of the photoemission spectra gave an estimate of $U$=7.0eV~\cite{Mizokawa}.
These studies 
suggest that $U\simeq 10-30t$ is reasonable, consistent with our results.
It is noteworthy that the gap in our calculation at $U=20t$
corresponds to 
$0.5 \sim 0.8$eV, which is in the same order of magnitude as the experimental value $\sim 0.24$eV in NaNiO$_2$~\cite{Molenda}.
We also note that 
the CO stabilization energy, which is estimated from the energy difference between the $\sqrt{3} \times \sqrt{3}$ COM phase and a para phase, is $\sim$ $0.05t$ for 
the present parameters $\gamma_{\rm br}$ and $\lambda_{\rm br}$: 
This result leads a rough estimate that the CO stabilization energy is $\sim$ 0.01 - 0.02 eV, which is in the same order of magnitude as the CO transition temperature observed in AgNiO$_2$ (365K~\cite{Wawrzynska,Wawrzynska2}).

Finally, we make a brief comment on 
the absence of any explicit ordering in LiNiO$_2$. 
Since LiNiO$_2$ is also a Mott insulator with a gap of 0.2eV~\cite{Kanno}, 
we suppose that the compound is in the strongly-correlated region 
similar to NaNiO$_2$.
In our results, 
there exists phase competition among three insulating phases 
with different spin and orbital patterns, 
the $\tau_z$-OF SF, $\tau_x$-OF SAF, and 
$\tau_y$-OF SF orderings. 
The competition brings about a frustration in the spin and orbital sectors 
in the vicinity of the phase boundaries. 
To argue the consequence of such frustration, 
we need to go beyond the present mean-field analysis; 
however, we can expect severe suppression of the orderings and 
some liquid-like or glassy behavior in the spin-orbital coupled system. 
Hence, one possibility is that LiNiO$_2$ is located in such competing regime. 
It is noteworthy that the competition is brought about 
by explicitly taking account of the electron-phonon couplings, 
which have not been considered in the previous effective model approaches~\cite{Mostovoy,Dare,Vernay}.
In addition to this intrinsic phase competition, extrinsic defects on Li sites may play an important role in the glassy behavior. 
Furthermore, a long-range strain effect might also play a role 
through the frustrating orbital and lattice sectors~\cite{PDF}.
It is interesting to take account of these factors explicitly, by extending our 
model and analysis.
We leave this problem 
for a future 
study. 

\section{Summary}
\label{sec:conclusion}

We have 
investigated the ground state of
the multi-orbital Hubbard model with electron-phonon couplings 
by the Hartree-Fock approximation and the adiabatic approximation, 
in order to elucidate the origin of 
various phases observed in $A$NiO$_2$ in a unified way. 
We found the $\sqrt{3}\times\sqrt{3}$ 
charge-ordered metallic phase in the weakly-correlated region and the 
orbital-ferro spin-ferro ordered insulating phase in the strongly-correlated region.
The $\sqrt{3}\times\sqrt{3}$ 
charge-ordered metallic phase is stabilized 
by a compromise between Coulomb repulsions and 
effective attractive interactions from the breathing-type electron-phonon coupling 
as well as the Hund's-rule coupling. 
The electronic state is half metallic; 
up-spin electrons are localized at the charge-rich sites, 
but down-spin electrons are extended and almost free.
On the other hand, 
the 
orbital-ferro spin-ferro ordered insulating phase is stabilized by the Jahn-Teller coupling under strong electron correlation, 
with a help by the Hund's-rule coupling in the competition with other orbital-ordered phases. 
These two phases are promising candidates for the low-$T$ phases in AgNiO$_2$ and NaNiO$_2$, respectively.
A possible origin of the quite different electron repulsion between AgNiO$_2$ and NaNiO$_2$ might be a screening effect from the cation and oxygen $p$ bands.
The puzzling glassy behavior in LiNiO$_2$ might be ascribed to 
the competition among different spin and orbital ordered states in the strongly-correlated region, which occurs under a substantial Jahn-Teller type electron-phonon coupling. 

\section*{Acknowledgements}
The authors thank 
M. Imada, 
S. Watanabe, and 
Y. Yamaji for 
fruitful discussions.
This work was supported by Grants-in-Aid for Scientific Research 
(No. 17071003, 17740244, 19014020, and 19052008), 
Global COE Program ``the Physical Sciences Frontier", the Next Generation Super Computing 
Project, and Nanoscience Program, from MEXT, Japan.

\end{document}